# Bandgap Lowering in Mixed Alloys of Cs$_2$Ag(Sb$_x$Bi$_{1-x}$)Br$_6$ Double Perovskite Thin Films

*Zewei Li,[1,†] Seán Kavanagh,[2,3,4,†] Mari Napari,[5] Robert G. Palgrave,[2] Mojtaba Abdi-Jalebi,[6] Zahra Andaji-Garmaroudi,[1] Daniel W. Davies,[2,4] Mikko Laitinen,[7] Jaakko Julin,[7] Richard H. Friend,[1] David O. Scanlon,[2,4,8] Aron Walsh,[3,9] and Robert L. Z. Hoye*[,3]*

[1] Cavendish Laboratory, Department of Physics, University of Cambridge, JJ Thomson Avenue, Cambridge CB3 0HE, UK

[2] Department of Chemistry, University College London, 20 Gordon Street, London WC1H 0AJ, UK

[3] Department of Materials, Imperial College London, Exhibition Road, London SW7 2AZ, UK

[4] Thomas Young Centre, University College London, Gower Street, London WC1E 6BT, UK

[5] Zepler Institute for Photonics and Nanoelectronics, University of Southampton, University Road, Southampton SO17 1BJ, UK

[6] Institute for Materials Discovery, University College London, Torrington Place, London WC1E 7JE, UK

[7] Department of Physics, University of Jyväskylä, P.O. Box 35, 40014 University of Jyväskylä, Finland

[8] Diamond Light Source Ltd., Diamond House, Harwell Science and Innovation Campus, Didcot, Oxfordshire OX11 0DE, UK

[9] Department of Materials Science and Engineering, Yonsei University, Seoul 120-749, South Korea

∗Correspondence to: r.hoye@imperial.ac.uk

ABSTRACT

Halide double perovskites have gained significant attention, owing to their composition of low-toxicity elements, stability in air and long charge-carrier lifetimes. However, most double perovskites, including $Cs_2AgBiBr_6$, have wide bandgaps, which limits photoconversion efficiencies. The bandgap can be reduced through alloying with $Sb^{3+}$, but Sb-rich alloys are difficult to synthesize due to the high formation energy of $Cs_2AgSbBr_6$, which itself has a wide bandgap. We develop a solution-based route to synthesize phase-pure $Cs_2Ag(Sb_xBi_{1-x})Br_6$ thin films, with the mixing parameter $x$ continuously varying over the entire composition range. We reveal that the mixed alloys ($x$ between 0.5 and 0.9) demonstrate smaller bandgaps (as low as 2.08 eV) than the pure Sb- (2.18 eV) and Bi-based (2.25 eV) compounds, with strong deviation from Vegard's law. Through in-depth computations, we propose that bandgap lowering arises from the Type II band alignment between $Cs_2AgBiBr_6$ and $Cs_2AgSbBr_6$. The energy mismatch between the Bi and Sb $s$ and $p$ atomic orbitals, coupled with their non-linear mixing, results in the alloys adopting a smaller bandgap than the pure compounds. Our work demonstrates an approach to achieve bandgap reduction and highlights that bandgap bowing may be found in other double perovskite alloys by pairing together materials forming a Type II band alignment.

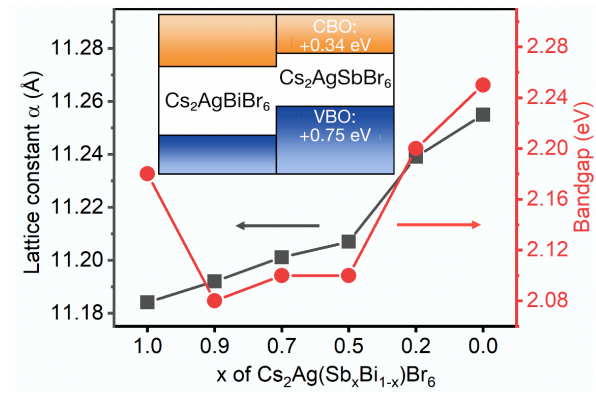

**KEYWORDS**: Double perovskites, lead-free, bandgap bowing, band alignment, solar absorbers, bandgap engineering, perovskite alloy, deformation potentials

MAIN TEXT

Despite the meteoric rise in efficiency of photovoltaics based on lead-halide perovskites, concerns remain over the toxicity of the lead component.[1] Three of the main approaches to find lead-free alternatives are: (a) chemical substitution of $Pb^{2+}$ with more benign metal cations;[2] (b) discovery of materials that could replicate the defect tolerance of the lead-halide perovskites;[3,4] and (c) exploration of halide materials that maintain the perovskite crystal structure. The third route has led to extensive work on double perovskites, in which the toxic $Pb^{2+}$ is replaced with a pair of monovalent and trivalent cations (*e.g.*, $Ag^+$ and $Bi^{3+}$).[5–11] Double perovskites, or elpasolites, have the general formula $A_2B(I)B(III)X_6$, where A is a monovalent cation (such as $Cs^+$ or $CH_3NH_3^+$), B(I) a monovalent metal cation (such as $Na^+$, $Cu^+$, $Ag^+$ or $Au^+$), B(III) a trivalent metal cation (usually $Sb^{3+}$, $Bi^{3+}$ or $In^{3+}$) and X a halide anion ($Cl^-$, $Br^-$ or $I^-$).[5,11–13] Calculations have shown that the bandgap can be tuned over a wide range by changing the species in each atomic site. There has been particular focus on $Cs_2AgBiBr_6$ for photovoltaic applications, and several groups have demonstrated long charge-carrier lifetimes, which can exceed a microsecond.[11,14,15] Photovoltaic devices with $Cs_2AgBiBr_6$ have demonstrated high external quantum efficiencies reaching 60%,[16] with stable performance over hundreds of hours in ambient air.[17] These promising charge transport properties, coupled with the improved stability of $Cs_2AgBiBr_6$ over lead-halide perovskites, have led to the material also being considered for photocatalytic applications.[18,19]

However, the potential of $Cs_2AgBiBr_6$ for photovoltaic and photocatalytic applications is limited by the wide bandgap (2.1–2.2 eV) and its indirect nature, which limit light absorption.[10,11,18–21] Efforts to achieve a smaller and more direct bandgap have involved alloying with Tl, but this is more toxic than Pb.[6,22] Alloying $Cs_2AgBiBr_6$ with the substantially less toxic Sb was also shown to be effective, but led to a smaller reduction in bandgap than Tl alloying.[23–25] In part, this was because the Sb-based double perovskite itself ($Cs_2AgSbBr_6$) has a wide bandgap of 1.9–2.1 eV.[9,13] Another factor was that only a maximum of 37.5% Sb could be introduced through powder melt synthesis, limiting the extent of bandgap reduction.[23] Intriguingly, mixed $Cs_2Ag(Sb_xBi_{1-x})Br_6$ alloys have exhibited a non-linear reduction in the bandgap, i.e., bandgap bowing, with increasing Sb content.[23,25] Bandgap bowing has also been found in Pb/Sn perovskite alloys, and, in this case, the bowing is pronounced, such that the mixed alloy exhibits a smaller bandgap than either of the pure Pb- or Sn-based compounds.[26–28] An important question would be whether a similar phenomenon occurs in $Cs_2Ag(Sb_xBi_{1-x})Br_6$ alloys.

So far, there is a lack of understanding of the extent of bandgap bowing in Sb-Bi double perovskite alloys and whether the mixed compositions could exhibit a smaller bandgap than the pure compounds. This requires the full composition range in $Cs_2Ag(Sb_xBi_{1-x})Br_6$ alloys to be synthesized. However, synthesizing phase-pure $Cs_2AgSbBr_6$ is challenging, particularly in thin film form. This is due to the high formation energy of $Cs_2AgSbBr_6$ and the small ionic radius of $Sb^{3+}$ relative to $Bi^{3+}$.[29] Recently, Liu *et al.* synthesized thin films of $Cs_2Ag(Sb_xBi_{1-x})Br_6$ alloys by dipping their substrates in a heated solution of the precursor salts dissolved in dimethyl sulfoxide (DMSO).[25] But they were only able to achieve up to 75% Sb alloying and were not able to synthesize the pure Sb-based compound. Colloidal nanocrystal synthesis routes have been found to be more successful in growing less thermodynamically-favored compounds (*e.g.*, iodide-based double perovskites, which have a positive heat of formation). Yang *et al.* recently developed a route to grow $Cs_2AgSbBr_6$ nanocrystals.[29] But it is also desirable to achieve phase-pure $Cs_2AgSbBr_6$ as bulk thin films, which do not have carriers confined within individual grains, as is the case for nanocrystals bound with long-chain ligands.

In this work, we achieved the synthesis of Sb-Bi double perovskite alloys over the entire composition range in thin film form. We found that the mixed alloys have a lower bandgap than the pure Bi- and Sb-based double perovskites, with similar pronounced nonlinear bandgap behavior to that found in lead-tin perovskite alloys.[26–28] Through X-ray diffraction (XRD) and Rutherford Backscattering Spectrometry (RBS), we found all films to be phase-pure, with the thin film composition in the bulk matching the ratio of elements in the precursor solution. We used Photothermal Deflection Spectroscopy (PDS) to accurately determine how the indirect bandgap and electronic disorder in Sb-Bi double perovskite alloys changed with composition. The electronic structure and band positions were calculated by Density Functional Theory (DFT) to provide insight into the origins of the bandgap lowering observed in the Sb-Bi double perovskite alloys.

$Cs_2Ag(Sb_xBi_{1-x})Br_6$ thin films were synthesized by solution processing. The CsBr, AgBr, $SbBr_3$ and $BiBr_3$ precursors were mixed according to their stoichiometric ratio in DMSO with a concentration of 0.5 mol·L$^{-1}$. In contrast to the dipping method used by Liu *et al.*,[25] we used a lower annealing temperature and shorter annealing time (detailed in the Experimental section in the SI), which may have contributed to the successful synthesis of phase-pure $Cs_2AgSbBr_6$. Photographs of the films (approx. 200 nm thickness in all cases) are shown in Fig. 1a, and it was observed that the mixed compositions have a deeper color. To determine the bulk composition of the films, we performed RBS measurements (Fig. 1b and Table 1). It was found that the Sb/Bi ratio from the RBS measurements matched the stoichiometry in the precursor solution, indicating no change in the composition during crystallization. This is indicative of no phase impurities precipitating out during film synthesis.

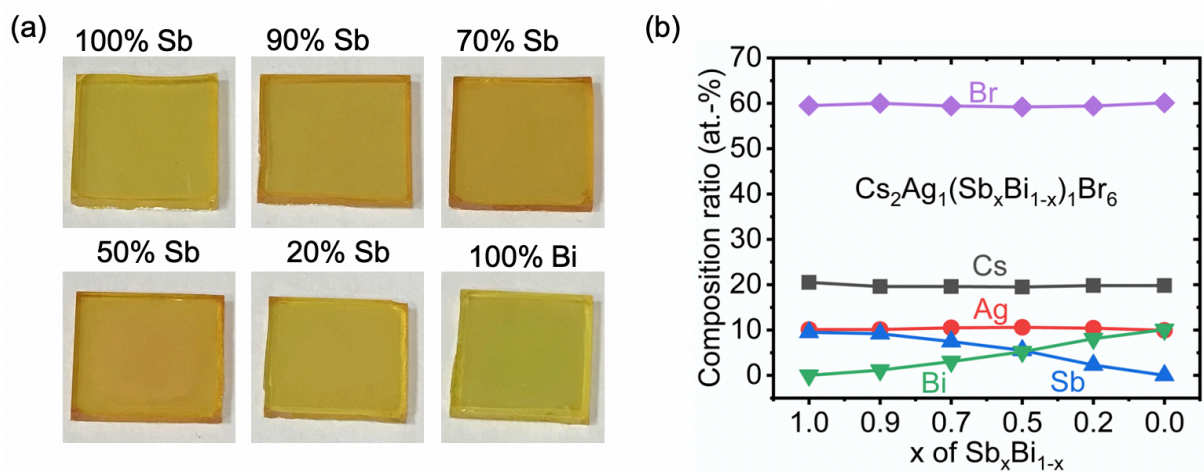

**Figure 1**. (a) Photographs of the $Cs_2Ag(Sb_xBi_{1-x})Br_6$ films on glass under ambient lighting, with a thickness of approximately 200 nm. The Sb percentage labelled is that in the precursor solution. (b) Measurement of the bulk composition of $Cs_2Ag(Sb_xBi_{1-x})Br_6$ films with Rutherford Backscattering Spectrometry (fitting in Fig. S1, SI)

**Table 1**. Quantification of the bulk composition of $Cs_2Ag(Sb_xBi_{1-x})Br_6$ thin films by RBS.

| Precursor Sb/(Sb+Bi) in solution | RBS | | | | | |
|---|---|---|---|---|---|---|
| | Cs/at.-% | Ag/at.-% | Sb/at.-% | Bi/at.-% | Br/at.-% | Sb/(Sb+Bi) in film |
| 100% | 20.5 | 10.1 | 9.5 | 0 | 59.5 | 100% |
| 90% | 19.6 | 10.1 | 9.2 | 1.2 | 60.0 | 88.5% |
| 70% | 19.6 | 10.5 | 7.5 | 3.2 | 59.4 | 70.1% |
| 50% | 19.5 | 10.6 | 5.5 | 5.2 | 59.2 | 51.4% |
| 20% | 19.8 | 10.4 | 2.3 | 8.1 | 59.4 | 22.1% |
| 0% | 19.8 | 10.0 | 0 | 10.1 | 60.1 | 0% |

The phase-purity of the films was determined through XRD measurements (Fig. 2a). $Cs_2AgSbBr_6$ and $Cs_2AgBiBr_6$ have the same cubic structure and space group (Fm$\bar{3}$m). The lattice constants were found to be 11.184 Å for $Cs_2AgSbBr_6$ and 11.255 Å for $Cs_2AgBiBr_6$, with a continuous variation between these extremes through the alloying range, as could be seen from the continuous shift in peak positions (Fig. 2b and Fig. 3d). Common phase impurities for the Sb-Bi double perovskites are $Cs_3Bi_2Br_9$ and $Cs_3Sb_2Br_9$. Both impurities have almost the same diffraction patterns with their double perovskite counterparts, but one of the main differences is a peak from $Cs_3Bi_2Br_9$ at 8.95°, or from $Cs_3Sb_2Br_9$ at 9.10°.[30] Neither impurity peaks were found to be present here. However, these XRD measurements were taken using a 1D linescan, and it is possible that the impurity peaks were not detected due to preferred orientation. We therefore performed 2D XRD measurements (Fig. S2, SI). We synthesized the whole alloying range of double perovskite thin films and intentionally induced a $Cs_3(Bi,Sb)_2Br_9$ phase impurity to the films with 50% Sb. From the 2D XRD patterns, we found that the diffraction peak should be detectable by the 1D linescan if it is present. Therefore, the absence of any phase-impurities in the 1D XRD patterns in Fig. 2a shows the films to be phase-pure.

The diffraction patterns also showed no peak splitting, but the full width at half maximum (FWHM) of the pure Sb-based double perovskite and mixed alloys were larger than the Bi-based double perovskite (Fig. 2b&c). This could either be due to smaller grains or small amounts of phase segregation that could not be resolved. Scanning Electron Microscopy (SEM) measurements of the films showed that the Sb-based double perovskites and mixed alloys have smaller grains than the pure Bi-based double perovskite (Fig. S3, SI), which is consistent with the trend in crystallite size calculated from the FWHM of the (022) diffraction peaks (Fig. 2c). These XRD and SEM measurements, together with the RBS measurements matching the stoichiometry of the precursor solutions, are consistent with the alloys being homogeneous.

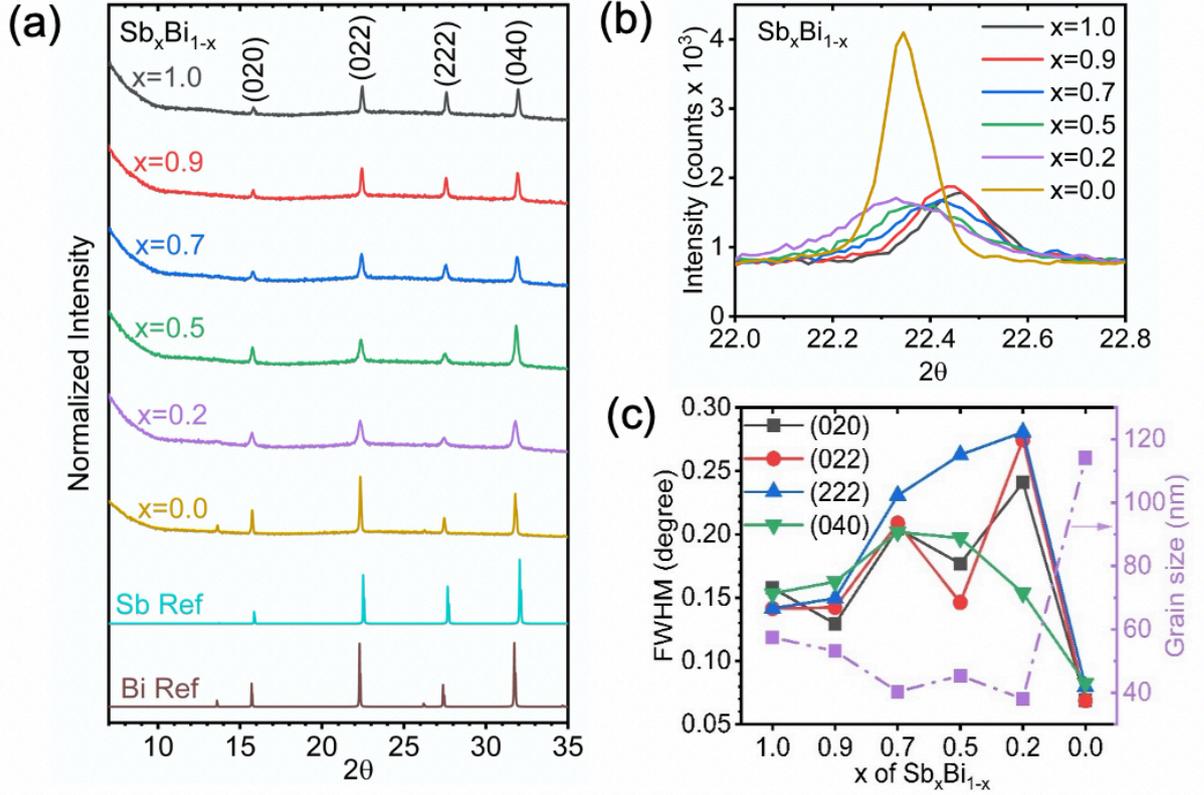

**Figure 2**. (a) XRD pattern of the $Cs_2Ag(Sb_xBi_{1-x})Br_6$ films over the composition series. (b) Full width half maximum (FWHM) of the (022) peak for different compositions. (c) FWHM of different diffraction peaks and the average grain size calculated using the Scherrer equation for different compositions

Standard optical transmittance and reflectance measurements (using a UV-visible spectrophotometer with an integrating sphere) were used to determine the absorption coefficient of the double perovskite thin films (Fig. 3a). All films showed a slow increase in the absorption coefficient for photon energies >2 eV, before rising sharply to >$10^5$ cm$^{-1}$ at >2.6 eV. An absorption coefficient exceeding $10^5$ cm$^{-1}$ at >2.6 eV is characteristic of a direct band-to-band transition. Both $Cs_2AgBiBr_6$ and $Cs_2AgSbBr_6$ have indirect bandgaps due to the effects of the Ag $d$ orbitals at the band-edges, and the alloys are also expected to have indirect bandgaps.[6] Therefore, to determine the indirect bandgap, we investigated the absorption onset in the lower photon energy range (*i.e.*, between 2–2.6 eV), leaving the discussion of the absorption in the higher photon energy range to later in the paper. We observed that the alloys with 70–90% Sb showed a lower-energy absorption onset than the pure Sb- and Bi-based films. However, the indirect bandgaps resulted in the absorption coefficients being small close to the band-edge, making it difficult to accurately determine the bandgap from standard transmittance and reflectance measurements. Therefore, we performed PDS measurements, which are sensitive to the absorbance 3–4 orders below the band-edge (operating details in the Experimental section, SI).

A consequence of the high sensitivity to low absorbance is that the measurements in PDS saturate for high absorbance at >2.6 eV (Fig. 3b), and the peaks in the absorption coefficient seen in Fig. 3a are not observed in the PDS measurements plotted on a semi-logarithmic scale (see later in the paper for a discussion of these peaks). The normalized absorbance from PDS measurements at the band edge (in the range of 2–2.3 eV) is approximately 3 orders of magnitude lower than the absorbance at 2.6 eV, where the absorption coefficient is >$10^5$ cm$^{-1}$ (Fig. 3a). Thus, the absorption coefficient at the band edge should be on the order of $10^2$ cm$^{-1}$. We note that silicon, which is a typical indirect bandgap material, has an absorption coefficient of $10^2$ cm$^{-1}$ at the band edge.[31,32]

The absorption plots shown in Fig. 3b do qualitatively indicate an exponentially increasing absorption coefficient versus energy, which appears linear on a semilogarithmic plot. The double perovskites with $x$ ranging from 0 to 0.7 have similar slopes in their absorption onsets when plotted on a semilogarithmic scale. For $x = 0.9$ and the pure Sb-based compound, the slopes are shallower. Increasing the fraction of Sb ($x$) from 0 (pure Bi) to 0.5 reveals a red-shift in the absorption onset. The double perovskites with $x = 0.7$ and $x = 0.9$ all have similar absorption onsets to the material with $x = 0.5$. These mixed alloys with $x = 0.5–0.9$ all have absorption onsets that are red-shifted to the pure Sb-based compound, as well as the pure Bi-based compound. Therefore, alloys with 50–90% Sb demonstrate smaller bandgaps than either of the pure end double perovskites.

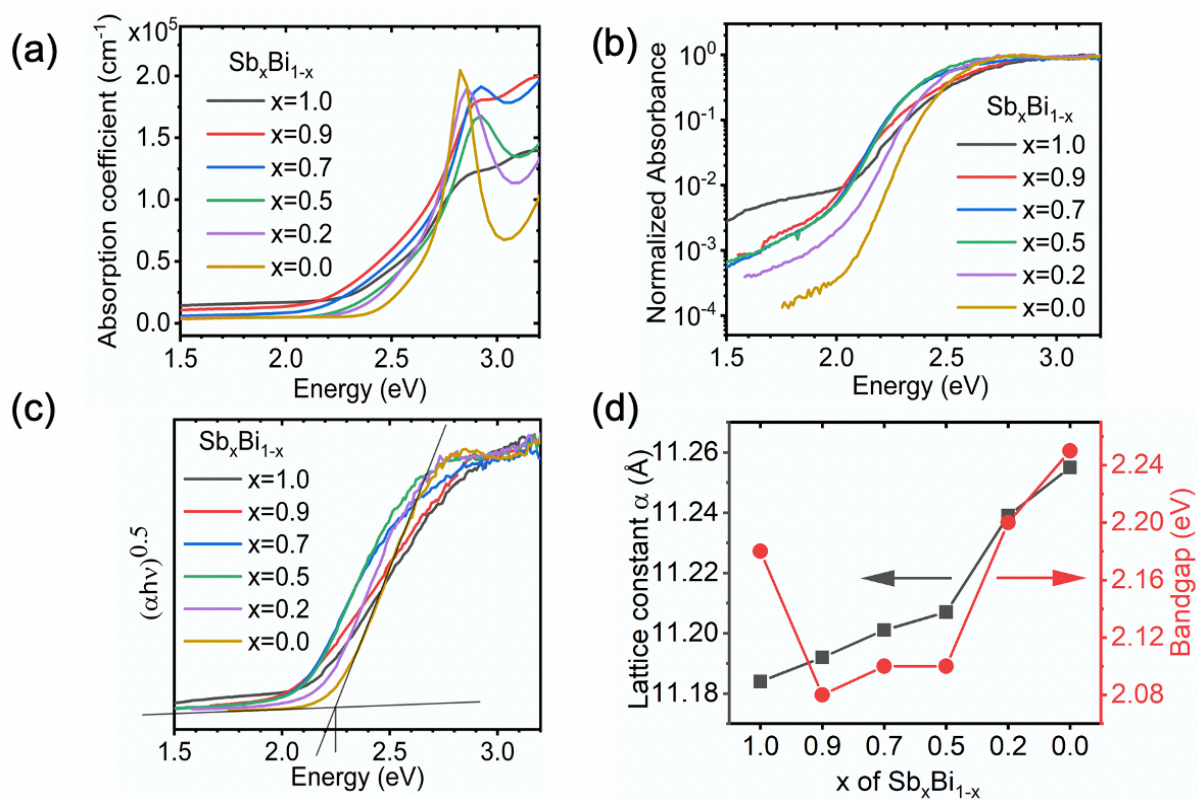

**Figure 3**. Measurements of the bandgap of $Cs_2Ag(Sb_xBi_{1-x})Br_6$ films. (a) Absorption coefficient determined from transmittance and reflectance measurements of films deposited on glass substrates. (b) Normalized absorbance measured by PDS for films deposited on quartz substrates. (c) Tauc plot from PDS measurements. (d) Bandgap and lattice parameter of double perovskite films plotted against the Sb content in the precursor solution. The lattice parameter was acquired by fitting the XRD measurements (Fig. 2a). The bandgap was obtained from the Tauc plot constructed from the PDS data in part c

To extract numerical values of the bandgap and quantify their variation with composition, we fitted the PDS absorbance data using a Tauc plot, which is a plot of $(\alpha h\nu)^{1/n}$ against $h\nu$. The rationale can be seen from Eq. 1

$$(\alpha h\nu)^{1/n} = A(h\nu - E_g) \qquad (1)$$

where $\alpha$ is the absorption coefficient, $h$ is Planck's constant, $\nu$ is the photon frequency, $A$ is a proportionality constant, $E_g$ is the bandgap, and $n = 2$ for an indirect bandgap.[31,32] Therefore, by plotting $(\alpha h\nu)^{0.5}$ against $(h\nu)$, the bandgap can be extracted from the intercept between the tangents fit to the absorption onset and background (Fig. 3c). From these Tauc plots, we found the system to exhibit significant bandgap bowing (Fig. 3d). The compound with $x = 0.9$ demonstrated the lowest bandgap of 2.08 eV, which is smaller than the bandgap of $Cs_2AgSbBr_6$ (2.18 eV) and $Cs_2AgBiBr_6$ (2.25 eV). It is possible that the lower bandgap extracted for the compound with $x = 0.9$ than for materials with $x = 0.5$ and $x = 0.7$ (both 2.10 eV) were due to errors in the fitting process, which may have arisen in part from the lower slope in the absorption onset of the material with $x = 0.9$. Nevertheless, we emphasize that from the Tauc plots, the compounds with $x = 0.5$–$0.9$ all have smaller bandgaps than the pure Bi- and Sb-based compounds, which is consistent with our analysis of the PDS absorbance measurements in Fig. 3b. This is also consistent with the trend in bandgaps we extracted from the Tauc plot (Fig. S4, SI) based on the absorption coefficient measurements (Fig. 3a). In addition, the bandgap lowering was smooth, with no switch in the bandgap from indirect to direct. This bandgap lowering behavior is similar to the Pb-Sn perovskite system and has not been previously observed in Sb-Bi double perovskites.

To understand the atomistic origin of the bandgap lowering in the Sb-Bi double perovskites, calculations of the electronic structure were performed. The band structure and density of states for pure Sb and Bi double perovskites are shown in Fig. 4a-d. In agreement with previous literature, as well as our optical measurements in Fig. 3, both $Cs_2AgSbBr_6$ and $Cs_2AgBiBr_6$ were found to have an indirect bandgap,[6,13,23] with the valence band maximum (VBM) located at the **X** point and the conduction band minimum (CBM) at the **L** point in reciprocal space.

Using relativistic hybrid density functional theory (HSE06 including spin-orbit coupling), the indirect bandgap and first direct transition of $Cs_2AgSbBr_6$ were calculated to be 1.37 and 2.60 eV, respectively. Analysis of the electron band orbital character revealed the conduction band states to be dominated by anti-bonding Sb 5p - Br 4p interactions, while the valence band states primarily arise from Sb 5s, Ag 4d and Br 4p orbital interactions. The $Cs^+$ cations act as spectator ions, as noted in literature,[33] with a relatively large nearest-neighbor-ion distance to $Br^-$.

Likewise, the indirect bandgap and direct transition of $Cs_2AgBiBr_6$ were calculated to be 1.77 and 2.49 eV, respectively. Due to the stronger spin-orbit coupling of the Bi 6$p$ orbitals, the direct bandgap of $Cs_2AgBiBr_6$ is slightly reduced from that of $Cs_2AgSbBr_6$, despite a larger fundamental gap, and is located at a different point in reciprocal space (**X** = (0.5, 0, 0.5), *c.f.* **b** = (0.5, 0.38, 0.62) for $Cs_2AgSbBr_6$). Similarly, the conduction band states are dominated by anti-bonding interactions between the trivalent pnictogen cation (Bi 6p) and the halide anion (Br 4p) orbitals, while the valence band states primarily arise from Bi 6s, Ag 4d and Br 4p orbital interactions. Notably, the larger band gap of $Cs_2AgBiBr_6$ relative to $Cs_2AgSbBr_6$ is an interesting exception to the typical trend of decreasing bandgap upon atomic substitution with heavier members from the same periodic group, which has been observed for other perovskite materials.[9,34,35]

To delve closer to the origin of bandgap lowering, the 'natural' band offset of the two double perovskite materials was calculated following the alignment procedure of Butler *et al.*[36], with the resulting electron band alignment diagram shown in Fig. 4e. Our investigations indicate a Type II 'staggered gap' alignment, with a valence band offset (VBO) of 0.75 eV and a conduction band offset (CBO) of 0.34 eV ($Cs_2AgSbBr_6$ above $Cs_2AgBiBr_6$ in both cases). Typically, the energies of valence electrons become less negative as one moves down a group in the periodic table, corresponding to a decrease in ionization energies. However, for $Bi^{3+}$ and Sb$^{3+}$, this is not the case, with the Bi$^{3+}$ 6s$^2$ lone pair being lower in energy than the Sb$^{3+}$ 5s$^2$ lone pair.[37] This is a result of the relativistic contraction of the Bi 6s orbital, due to its large atomic number, resulting in a more localized, lower-energy valence s orbital, compared to Sb 5s.[38] Consequently, the more-diffuse, higher-energy Sb 5s$^2$ lone-pair has a stronger interaction with the Ag 4d and Br 4p orbitals, due to a reduced energy separation of the bonding orbitals (Fig. S5, SI). This produces both greater dispersion in the valence band and a higher VBM, as shown in Fig. 4e. In a similar manner, the elevated position of the CBM in Cs$_2$AgSbBr$_6$ can be understood by considering the ionization energies and ionic orbital energies. While the first ionization energy of Sb is indeed larger than that of Bi, the third ionization energy (corresponding to the removal of a p electron from Sb$^{2+}$/Bi$^{2+}$) is in fact 0.3 eV lower for Sb than for Bi,[37] indicating higher energy cationic p states. When the pnictogen elements are in the +3 oxidation state, as is the case in these materials, the third ionization energy provides an improved estimate for the energy of the unoccupied p orbitals. Hence, we argue that, in the +3 formal charge state, the Sb valence p orbitals are in fact higher in energy than those of Bi, suggesting a higher energy CBM for Cs$_2$AgSbBr$_6$, as witnessed in our investigations.

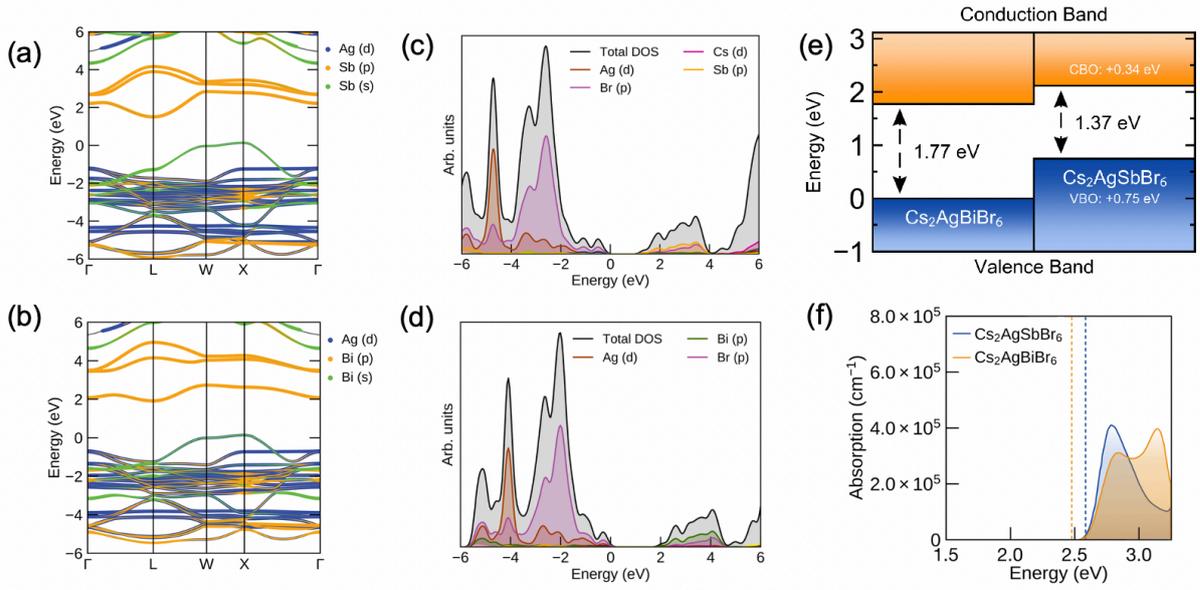

**Figure 4**. (a) Orbital-projected band structure of $Cs_2AgSbBr_6$. (b) Orbital-projected band structure of $Cs_2AgBiBr_6$. (c) Electronic density of states of $Cs_2AgSbBr_6$. (d) Electronic density of states of $Cs_2AgBiBr_6$. (e) Electron band alignment of $Cs_2AgSbBr_6$ and $Cs_2AgBiBr_6$. CBO = Conduction Band Offset, VBO = Valence Band Offset. (f) Calculated optical absorption plots of $Cs_2AgSbBr_6$ (blue) and $Cs_2AgBiBr_6$ (orange). Dashed vertical lines indicate the positions of optical absorption onset, corresponding to the direct band transitions of 2.60 eV for $Cs_2AgSbBr_6$ and 2.49 eV for $Cs_2AgBiBr_6$. Note that only vertical transitions are accounted for in this calculation, and the absorption from indirect transitions are not shown.

The origins of bandgap bowing in semiconductor alloys are typically due to chemical effects (*e.g.*, differences in electronegativity), local size-mismatch effects or changes in the lattice parameter.[39–41] For the Pb-Sn perovskite alloy system, Im *et al*. attributed bandgap bowing to the combined effect of spin-orbit coupling and composition-induced phase change.[28] Snaith and coworkers proposed that the short range ordering of preferred atomic scale clusters allow the bandgap of the mixed alloy to be below that of both pure compoounds.[27] In contrast, Stevanović and coworkers found that the strong nonlinearity in bandgap was primarily due to the mismatch in energy between s and p atomic orbitals of Pb and Sn.[26] The Pb-Sn alloys have a higher VBM dominated by Sn-5s and I-5p orbitals and lower CBM dominated by Pb-6p and I-5p orbitals. They found that spin-orbit coupling, structure changes and short-range ordering did not have a significant effect on bandgap bowing, and proposed that a homogeneous structure would have improved stability over a structure with short range ordering.

In the case of $Cs_2Ag(Sb_xBi_{1-x})Br_6$, we propose that the Type II staggered gap alignment between the pure compounds allows for the non-linear mixing of electronic states such that bandgaps lower than that of either pure material are obtained (Fig. 4e). Upon addition of Bi to the pure Sb double perovskite, the conduction band will be lowered because the CBM wavefunction amplitude is preferentially allocated to the Bi sites, producing a 'Bi-like', lower-energy conduction band state. The presence of Bi will also enhance spin-orbit coupling, further reducing the energy of the CBM. On the other hand, the highest energy valence band state will remain 'Sb-like', with greater wavefunction amplitude at the Sb sites yielding a VBM only slightly below that of the pure Sb material. This mixing of electronic states in the alloys to produce a low-energy CBM, dominated by Bi-Br interactions, and a high-energy VBM, dominated by Sb-Ag-Br interactions, produces bandgaps in the double perovskite alloys which are lower than that of either pure material.

As the alloy mixing parameter $x$ approaches extreme values ($x \to 0$ or $x \to 1$), the ability of the band extrema states to simultaneously adopt Bi and Sb character is diminished, hence the bandgap increases toward the pure double perovskite values. Therefore, we propose that this orbital-mixing behavior, facilitated by the Type II bandgap alignment, is the origin of the non-linear, non-monotonic variation in bandgap with composition in the Sb-Bi double perovskite system.

Another possible contribution to bandgap bowing is volume deformation, whereby variation in the lattice constant upon alloying results in non-linear transformation of the electronic structure.[26] This mechanism, however, more commonly dominates in alloys involving more chemically-distinct materials than is the case here. Moreover, due to the small positive bandgap deformation potentials calculated for the Sb and Bi compounds ($\Delta E_g \sim 0.02$ eV), we rule out this mechanism. Relative to the experimentally-observed bowing (~ 0.1 eV, Fig. 3d), volume distortion alone is not the origin of bandgap bowing in this alloy system.

Finally, we return to discuss the sharp peak in the absorption coefficient of the films at >2.6 eV (Fig. 3a). This is especially evident for $Cs_2AgBiBr_6$. Previous work attributed the sharp absorption peak in $Cs_2AgBiBr_6$ to an exciton associated with the direct transition.[24,42–45] However, there is also the possibility that these features in the pure compounds and mixed alloys are due to a narrow density of states at the band-edges. To explore this possibility, we computed the absorption spectra for $Cs_2AgSbBr_6$ and $Cs_2AgBiBr_6$ without excitonic effects (Fig. 4f). These spectra were broadened by convoluting with a Gaussian peak with a FWHM of 0.15 eV. For both materials, the calculated absorption spectra exhibit peaks at 2.8 eV, in agreement with the experimental measurements. These peaks arise from the relatively-weak dispersion of the electron bands at the CBM and VBM of the materials (Fig. 4a, b), yielding peaks in both the density of states (Fig. 4c, d) and thus the optical absorption (Fig. 4f). This strongly suggests that the peaks observed in the experimental UV-visible measurements are the result of direct transitions between the relatively-flat electron bands. That said, the exciton binding energy of $Cs_2AgBiBr_6$ was calculated to be 167 meV within effective mass theory, which is sufficiently large that stable exciton formation is possible in these systems.

Curiously, the computed absorption spectrum for $Cs_2AgSbBr_6$ shows a distinct absorption peak (Fig. 4f), whereas the measured peak from UV-visible spectrophotometry became less distinct with increasing Sb content, until there was barely an observable peak for the pure Sb-based compound (Fig. 3a). This may have been due a 'smearing out' of the absorption peak due to structural disorder.[46–48] We calculated the absorption spectra for the pure Sb-based compound convoluted with Gaussian peaks with wider FWHM (0.2 eV, 0.3 eV, 0.4 eV; Fig. S6, SI). These show that the peak becomes indistinguishable when broadening is large, as witnessed experimentally and consistent with the PDS measurements. Thus, peak broadening as a result of disorder leads to the observed 'smeared-out' absorption spectrum in Fig. 3a.

In conclusion, we successfully synthesized phase-pure $Cs_2AgSbBr_6$ thin films, as well as $Cs_2Ag(Sb_xBi_{1-x})Br_6$ with $x$ varying over the full compositional range. In doing so, we found that the mixed double perovskites with $x$ between 0.5 and 0.9 to have the smallest bandgaps, lower than those of the pure compounds. From the electronic band alignment, we found that the origin of bandgap bowing in this double perovskite alloy is due to chemical rather than structural effects. The Type II band alignment between $Cs_2AgBiBr_6$ and $Cs_2AgSbBr_6$, in combination with non-linear mixing of the electronic states, results in the alloy having smaller bandgaps than either pure material. Our work demonstrates a novel route to reduce the bandgap of $Cs_2AgBiBr_6$ and $Cs_2AgSbBr_6$, which could be generalized to other halide double perovskites. That is, we propose that alloys formed from compounds with a Type II band alignment could exhibit similar bandgap lowering. This may prove crucial for improving the suitability of double perovskites for photovoltaic and photocatalytic applications.

ASSOCIATED CONTENT

**Supporting information**

The Supporting Information is available free of charge on the ACS Publications website at DOI: XXXX.

Supporting information includes the following parts: film deposition methods, computation methods, characterization methods, fitting of the RBS measurements, 2D XRD patterns, SEM of $Cs_2Ag(Sb_xBi_{1-x})Br_6$ thin films, Tauc plot fitting of the UV-visible spectroscopy data, schematic molecular orbital diagram of $Cs_2AgSbBr_6$ and $Cs_2AgBiBr_6$, and calculated optical absorption spectra.

AUTHOR INFORMATIONS

**Author contribution**

†Z.L. and S.K. contributed equally to this work.

Z.L. synthesized the Sb-Bi double perovskite thin films, took photographs, and performed 1D XRD, SEM and UV-Visible Spectrophotometry measurements. S.K. performed the theoretical calculations and analyzed the results, along with D.D, D.O.S. and A.W. M.N. analyzed the RBS results, which were collected by M.L. and J.J. R.G.P. measured the 2D XRD data. M.A.-J. and Z.A.-G. measured and analyzed the PDS data. R.H.F. supervised Z.L. and contributed to the discussion. R.L.Z.H. conceived of the project and supervised the work. All authors discussed the results and contributed to writing the paper.

**Corresponding author**

∗Email: r.hoye@imperial.ac.uk

**Notes**

The authors declare no competing financial interest.

ACKNOWLEDGEMENTS

Z.L. would like to thank Cambridge Trust and Chinese Scholarship Council for financial support. S.K. acknowledges the use of the UCL Grace High Performance Computing Facility (Grace@UCL), the Imperial College Research Computing Service ([doi.org/10.14469/hpc/2232](doi.org/10.14469/hpc/2232)), and associated support services, in the completion of this work. Via membership of the UK's HEC Materials Chemistry Consortium, which is funded by EPSRC (EP/L000202), this work also used the UK Materials and Molecular Modelling (MMM) Hub for computational resources, which is partially funded by the EPSRC (EP/P020194). R.L.Z.H. acknowledges support from the Royal Academy of Engineering under the Research Fellowship programme (No. RF\201718\1701), the Isaac Newton Trust (Minute 19.07(d)), and the Kim and Juliana Silverman Research Fellowship at Downing College, Cambridge. M.A.-J. thanks Cambridge Materials Limited, Wolfson College, University of Cambridge and EPSRC (grant no. EP/M005143/1) for their funding and technical support. M.L. and J.J. acknowledge the support by the RADIATE project under the Grant Agreement 824096 from the EU Research and Innovation programme HORIZON 2020.